\documentclass[final]{svjour2}
\usepackage{graphicx}
\usepackage{rotating}
\usepackage{amssymb}
\usepackage{mathptmx}
\usepackage[numbers]{natbib}
\makeatletter
\journalname{Journal of Low Temperature Physics}

\bibpunct{}{}{,}{s}{}{,}

\begin{document}

\newcommand{\hdblarrow}{H\makebox[0.9ex][l]{$\downdownarrows$}-}
\title{Low-Mass WIMP Sensitivity and  Statistical Discrimination of Electron and Nuclear Recoils by Varying Luke-Neganov Phonon Gain in Semiconductor Detectors}

\author{M. Pyle$^1$ \and D.A. Bauer$^2$ \and B. Cabrera $^1$ \and J. Hall$^2$ \and R.W. Schnee$^4$ \and R. Basu Thakur$^{2,3}$ \and S. Yellin$^1$}

\institute{1:Department of Physics, Stanford University, Stanford, Ca 94110, USA \\
                2:Fermi National Accelerator Laboratory,  Batavia, IL 60510, USA\\
                3: Department of Physics, University of Illinois at Urbana-Champaign, Urbana, IL 61801, USA\\
                4:Department of Physics, Syracuse University, Syracuse, NY 13244, USA\\
                \\
\email{mpyle1@stanford.edu}
}

\date{\today}

\maketitle


\begin{abstract}
Amplifying the phonon signal in a semiconductor dark matter detector can be accomplished by operating at high voltage bias and converting the electrostatic potential energy into Luke-Neganov phonons. This amplification method has been validated at up to $|E|=40$\,V/cm without producing leakage in CDMSII Ge detectors, allowing sensitivity to a benchmark WIMP with mass $M_{\chi}=8$\,GeV/c$^{2}$ and  $\sigma=1.8\times10^{-42}$\,cm$^{2}$ (with significant sensitivity for  $M_{\chi} > 2$\,GeV/c$^{2}$)  assuming flat electronic recoil backgrounds near threshold.  Furthermore, for the first time we show that differences in Luke-Neganov gain for nuclear and electronic recoils can be used to discriminate statistically between low-energy background and a hypothetical WIMP signal by operating at two distinct voltage biases. Specifically, 99\% of events have p-value $<  10^{-8}$ for a simulated 20\,kg-day experiment with a benchmark WIMP signal with $M_{\chi}=8$\,GeV/c$^{2}$ and $\sigma=3.3\times10^{-41}$\,cm$^{2}$.
\end{abstract}

\section{Motivation}
For low-mass WIMPs ($M_{\chi}<10$\,GeV/c$^{2}$), the energy transfer in an elastic WIMP-nucleon interaction is barely above detection threshold for dark matter experiments. In particular, at these recoil energies experiments may lose the ability to distinguish between electron recoils and nuclear recoils, have trouble defining a fiducial volume, and have poor trigger efficiencies. For these reasons, the limits at low masses are many orders of magnitude larger than the limits for a heavier  WIMP ($M_{\chi} \sim100$\,GeV/c$^{2}$). 

The CoGeNT\cite{CoGeNT} and DAMA\cite{DAMA} experiments have unexplained low-energy signals/modulation that are close enough to the CDMS \cite{CDMS-LE}  and XENON10\cite{Xenon10-LE} limits that some members of the dark matter physics community believe a misestimation of systematics could explain the experimental inconsistencies. The new  interdigitated CDMS detectors\cite{iZIP} should improve our sensitivity to low-mass WIMPs by an order of magnitude due to improved definition of the phonon fiducial volume at low energies, but even further improvement should be possible by magnifying their phonon signal for very low-energy events, an idea first discussed and studied by Paul Luke\cite{Luke8890}.

\section{Luke-Neganov Phonon Gain}

In a CDMS detector, the $e^{-}/h^{+}$ pairs produced in an interaction are drifted across the crystal to charge-amplifier-instrumented electrodes. While drifting, the carriers shed Luke-Neganov phonons equal in energy to the external electronic potential energy across the detector.  Thus, the total phonon energy, $P_{t}$, created in an interaction and measured with TES arrays is a sum of the phonons produced directly in the recoil itself ($P_{r}$) and the Luke phonons 
\begin{equation}
	P_{t}= P_{r} + n_{e/h} e V_{b}
	\label{eq-defpt}
\end{equation}
where $n_{e/h}$ is the number of  $e^{-}/h^{+}$ pairs produced and $V_{b}$ is the external bias. $n_{e/h}$ is a function of both $P_{r}$ and interaction type (nuclear/ electronic) and traditionally is written in terms of the average energy needed to produce an $e^{-}/h^{+}$ pair for an electronic recoil , $\epsilon_{e/h}$, and a normalized yield factor, $Y$, which by definition is 1 for electronic recoils and for nuclear recoils has been found to roughly follow Lindhard\cite{Lindhard} theory in Ge: 
\begin{equation}
	n_{e/h}= Y(P_{r},type) P_{r}/ \epsilon_{eh}   
\end{equation}
or 
\begin{equation}
	P_{t}= (1+ Y(P_{r},type)V_{b}/\epsilon_{eh})P_{r} .
\end{equation}

In standard operating mode CDMS measures both $P_{t}$ and $n_{e/h}$, and can thus estimate $Y$ for each particle interaction individually, allowing discrimination between electron and nuclear recoils.  To maximize this discrimination, we usually operate at small $V_{b}$ to minimize correlation between $P_{t}$ and $n_{e/h}$.  Paul Luke proposed that if instead we operate at large $V_{b}$, then the Luke phonon signal will completely dominate the intrinsic recoil signal and $P_{t}$ will simply be a calorimetric measure of $n_{eh}$ with gain proportional to $V_{b}$.  

\begin{figure}[h!]
\begin{center}
\includegraphics[ width=0.49\linewidth, keepaspectratio]{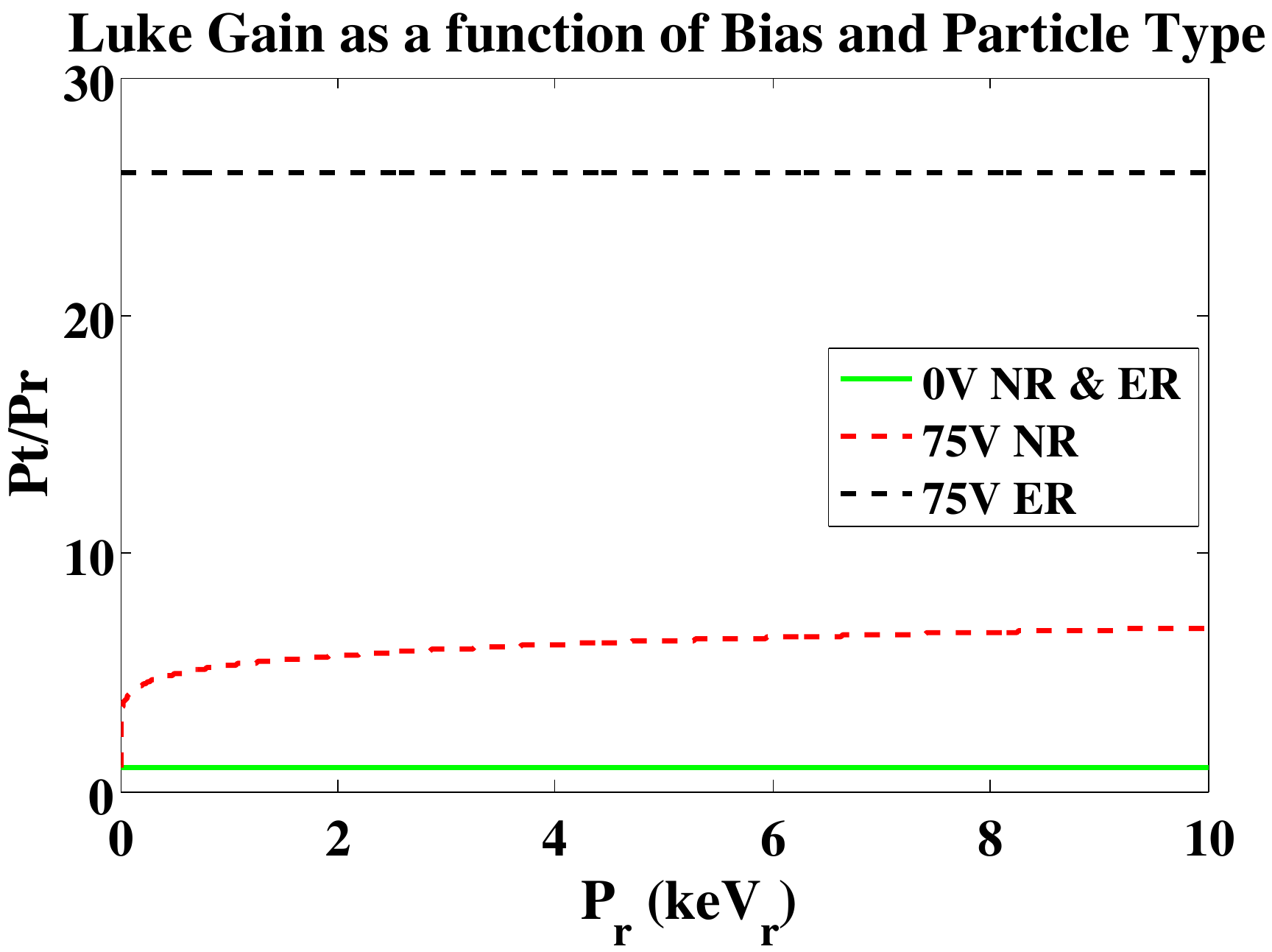}
\includegraphics[ width=0.49\linewidth, keepaspectratio]{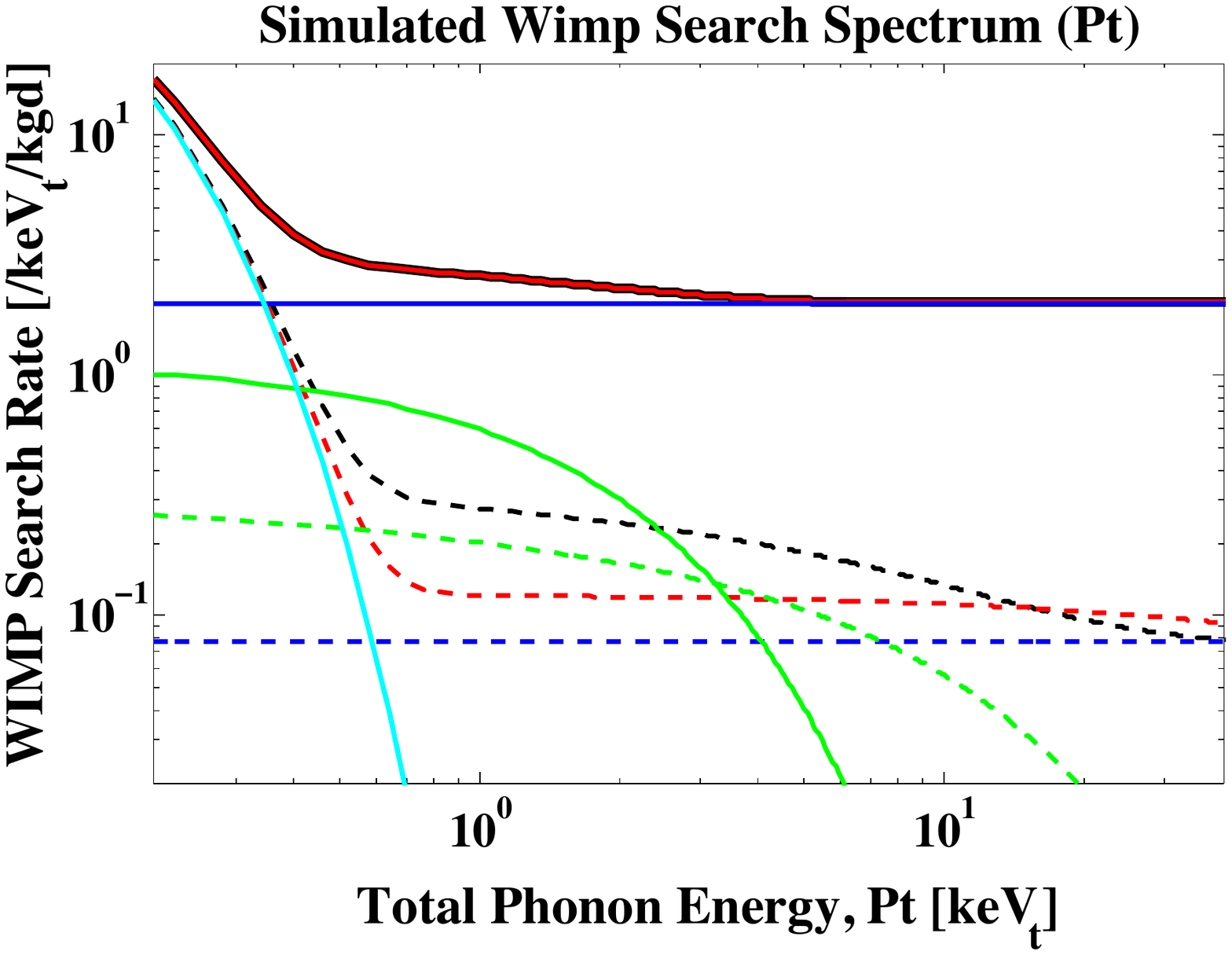}
\end{center}
\caption{(Color online) left:Luke-Neganov Gain for various $V_{b}$ and recoil type. right: total phonon rate estimates for 2 hypothetical signals. At 0\,V bias (solid curves), an exponentially increasing low energy plus constant electronic background (red) behaves identically to the sum (black) of a WIMP signal (green) and a flat electronic background (blue). At 75\,V,  these same two cases (same colors but dashed) are easily distinguishable. All cases include contributions from noise-induced triggers (cyan).}
\label{fig-LukeGain/rdf_pt}
\end{figure}

Unfortunately, there are two experimental constraints which limit the usefulness of Luke-Neganov gain. First, field and sensor geometries must be chosen to limit interaction between free carriers and crystal surfaces\cite{Cresst}. Second, at very high voltages (for CDMSII $|E| > \sim 40$\,V/cm) ambient current leakage dominates sensor johnson/TFN noise and thus for our 1" Ge devices, $V_{b}$ = 75V seems to be near optimum.  At this $V_{b}$,  the Luke-Neganov gain, $P_{t}/P_{r}$, for electron recoils is $\mathrm{\sim26}$ as shown in Fig.~\ref{fig-LukeGain/rdf_pt}. Consequently,  the threshold at which we expect good fiducial-volume definition in the  interdigitated CDMS detectors, $P_{t} > 750$\,eV, corresponds to electron recoil energy $E_R > 30$\,eV$_{ee}$. Of course, when operating in this mode event-by-event electron/neutron recoil discrimination is impossible.

Due to the dependence on yield of the Luke-Neganov gain, for nuclear recoils this gain factor is suppressed by  $\sim \times$5 in the limit of large $V_{b}$.  Unfortunately, this means that the fiducial volume threshold recoil energy corresponds to a nuclear-recoil energy $E_R > 150$\,eV$_{\mathrm{ne}}$, resulting in significant degradation of sensitivity for WIMP mass $M_{\chi} < 2$\,GeV/c$^{2}$.

On the bright side, this variation in Luke gain with particle type means that an electron recoil background is preferentially pushed to higher energies, energies which are not sensitive to a low mass WIMP signal. This effect is clearly seen in Fig.~\ref{fig-LukeGain/rdf_pt} where the black lines  correspond to a benchmark total recoil rate spectrum on a detector for a $M_{\chi} = 8$\,GeV/c$^{2}$ and $\sigma_{SI}= 5.0\times 10^{-42}$\,cm$^{2}$ WIMP (green) plus flat electron-recoil background of  2\,keV$_{\mathrm{r}}^{-1}$kg$^{-1}$day$^{-1}$ (blue) which is approximately the experimentally measured low-energy electronic background of the CDMS~II experiment\cite{CDMS-ER}. At  $V_{b}=0$\,V (solid lines), the benchmark WIMP interaction rate is dominated at all energies by background.  By contrast, at $V_{b}=75$\,V, the flat electron recoil background is relatively suppressed to the point that  for $P_{t}<4$\,keV$_{\mathrm{t}}$ the WIMP signal dominates.

This natural electronic background suppression means that the sensitivity of a $V_{b}=75$\,V Ge WIMP search experiment without background subtraction will be $\sigma_{SI} \sim 1.8\times 10^{-42}$\,cm$^{2}$ for $M_{\chi} = 8$\,GeV/c$^{2}$,  $\sim$50$\times$ lower than the CoGeNT signal region, \textbf{assuming no exponential increase in the electron recoil background at low energy}.

\section{Nuclear Recoil/Electron Recoil Statistical Subtraction}
The recoil-dependent variation in gain response also leads in Fig.~\ref{fig-LukeGain/rdf_pt} to visible shifts in spectrum shape. At 75\,V, the benchmark WIMP+background model (black) has a significant excess of events at low energies and a suppression of events at high energies relative to the exponential electronic background spectrum (red) which was chosen to precisely mimic 0\,V response. This behavior is general; \textbf{one can discriminate between electronic and nuclear recoil spectra by  measuring total phonon distributions at multiple voltages}. This possibility to differentiate between electronic background and low-mass WIMP signals is a powerful feature not shared by CoGeNT, DAMA, or S2-only XENON10 results and adds significant discovery potential to a CDMS experiment in this WIMP mass range.

\begin{figure}[t]
\begin{center}
\includegraphics[ width=0.55\linewidth, keepaspectratio]{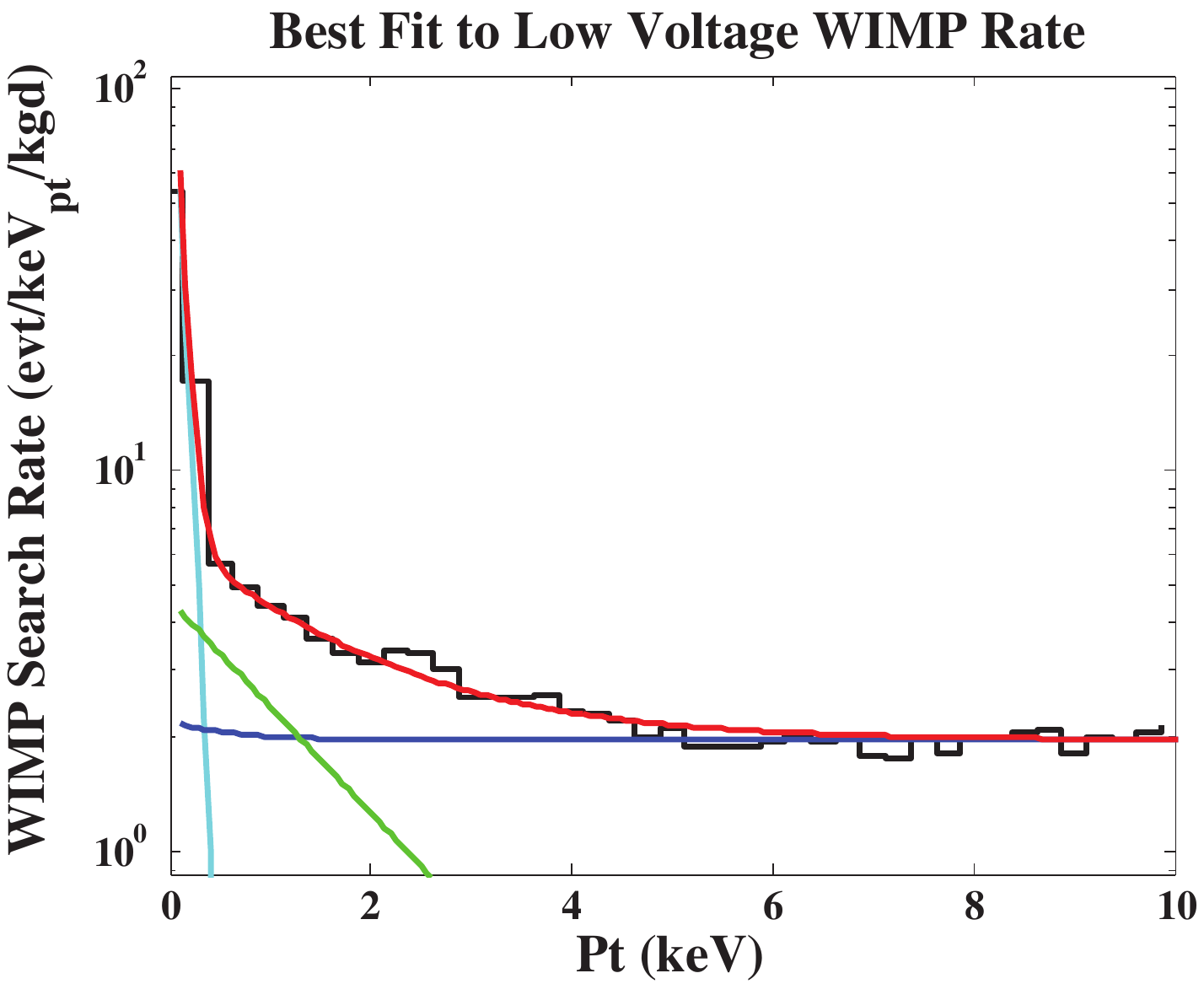}
\end{center}
\caption{(Color online) Fit of simulated WIMP+background spectrum (black) to the sum (red) of an exponential NR spectrum (green), a variable trigger threshold (cyan), and an exponentially decaying + constant electronic background (blue).}
\label{fig-ptfit}
\end{figure}

To quantify our discrimination capability, we simulate 1000 experiments detecting the previously mentioned benchmark rate for 400\,kg-days, and also simulate for a much larger WIMP rate, $\sigma_{SI}= 3.3\times10^{-41}$\,cm$^{2}$ for 20\,kg-days (benchmark 2).  To contrast with these signal simulations, we also simulate 1000 experiments with an electronic exponential background that is identically distributed for $V_{b}=$0\,V to maximally probe electronic/nuclear statistical discrimination. 
 We then fit each of these simulated rate measurements to a sum including a random noise contribution of the form $Ne^{-P_t/\epsilon_e}$, a contribution from electron recoils whose recoil energy is distributed according to form $C_e + R_e e^{-E_R/\epsilon_r}$, and, if WIMPs are included in the fit, a contribution from nuclear recoils with $E_R$ distributed according to $R_n e^{-E_R/\epsilon_n}$. The random noise is independent of $V_b$, the electron-recoil background spectrum is especially sensitive to $V_b$, and the nuclear-recoil spectrum is intermediate in sensitivity to $V_b$.  All contributions are constrained to be positive.  Fits without a nuclear-recoil contribution have 5 degrees of freedom (DOF); fits with a nuclear-recoil contribution have 7 DOF.
A fit of simulated data under the signal hypothesis is shown in Fig.~\ref{fig-ptfit}.

\begin{figure}[!ht]
\begin{center}
\includegraphics[ width=0.49\linewidth, keepaspectratio]{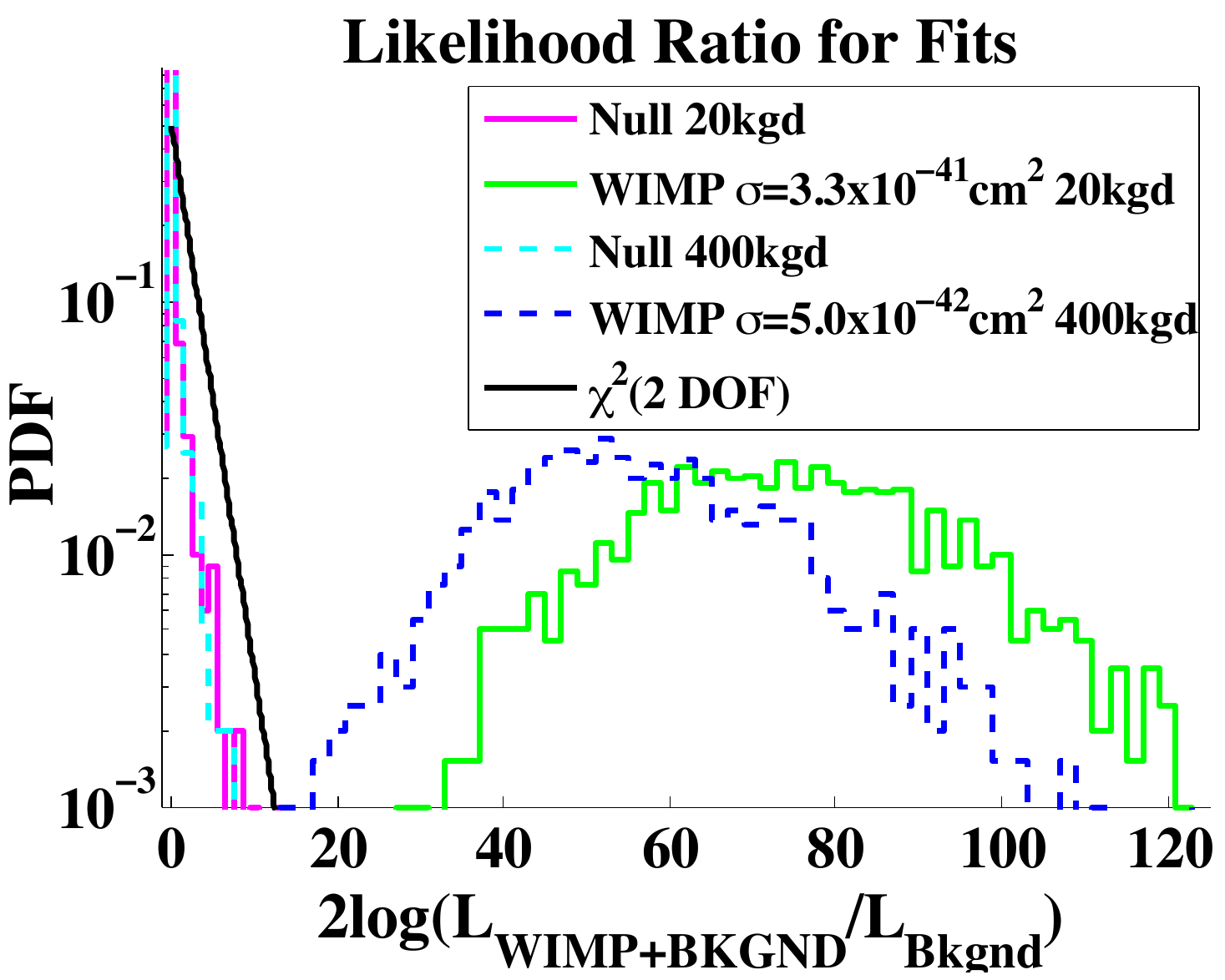}
\includegraphics[ width=0.49\linewidth, keepaspectratio]{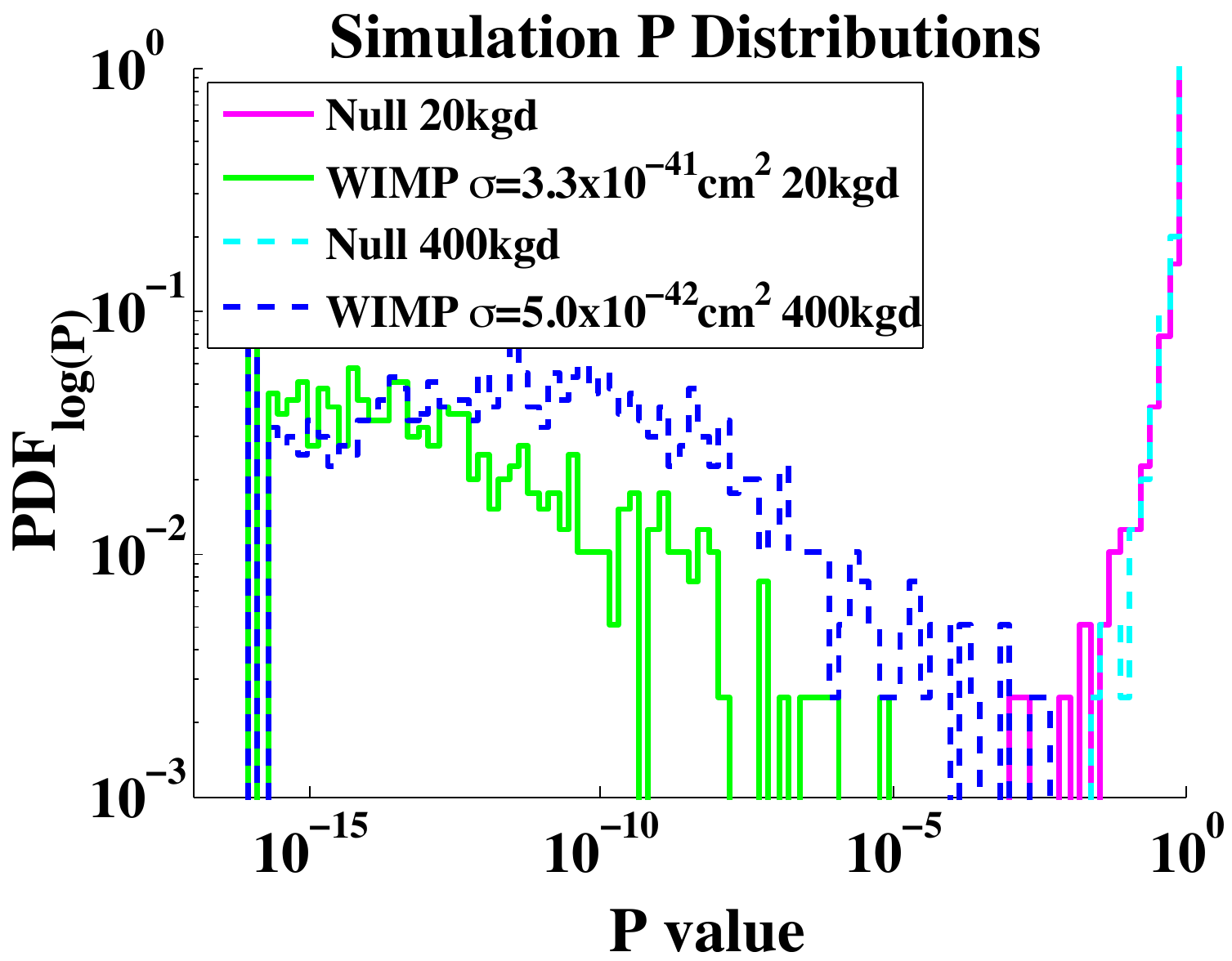}
\end{center}
\caption{(Color online) left: Log likelihood ratio for the signal+ background fit versus background only fit for the 2 WIMP model benchmarks (blue/cyan) and for the background only models (cyan/magenta). right: Statistical separability assuming a $\chi^{2}$ distribution with 2 DOF between null simulations (cyan/magenta) and simulations with WIMP signals (green/blue).}
\label{fig-fit_DlogL/Pval}
\end{figure}

For the background-only simulated experiments (cyan and magenta in Fig.~\ref{fig-fit_DlogL/Pval}), the fit quality for the 2 fits are statistically identical since the log of the likelihood ratio distribution are smaller than a $\chi^{2}$ distribution with 2 DOF (the physical constraints of positive rates means that the additional signal DOF are not always optimally used).  

By contrast, both signal simulations (blue/green) show sizeable fit quality differences, so much so that to the level of statistics simulated ($>$99.8\%), there is no overlap with the null simulations. We can further quantify the electron recoil/ nuclear recoil discrimination potential by looking at conservative p-values assuming a $\chi^{2}$ with 2 DOF  null distribution (fig. \ref{fig-fit_DlogL/Pval}). For the benchmarks, 99\% of simulated experiments have  p-values $< 10^{-8}$ and $<$ 2.4x$10^{-6}$ respectively, indicating very strong discrimination capability.  Best fit WIMP cross sections were found to be slightly systematically suppressed at $\mathrm{ 5x10^{-42} \pm 8x10^{-43}cm^{2} }$(90\%CL) and $\mathrm{ 3.3x10^{-41} \pm 3x10^{-42}cm^{2} }$(90\%CL)  for the two benchmarks.

\section{Potential Systematics}
With such large statistical discrimination capability, systematics will almost certainly dominate our final WIMP sensitivity. In order of importance, the dominant expected systematics include
 \begin{itemize}
 	\item Fiducial volume leakage: high-radius electronic recoils can have carrier trapping on the outer cylindrical surface that suppresses the Luke phonon gain and thus mimics a WIMP signal. Our standard collection and analysis of $\sim10^{6}$ $^{133}$Ba calibration events should be sufficient to allow estimation of  this effect and correction for it.   
	\item Fiducial volume variation with energy and voltage: misestimates of fiducial volume that artificially mimic rate distribution changes seen in fig. \ref{fig-LukeGain/rdf_pt} can produce a false WIMP signal but should be preventable with $^{252}$Cf calibration.
	\item Sensitivity to experimentally unverified Yield at ultra-low energies: through use of both $^{133}$Ba and  $^{252}$Cf (which in the CDMS II detector produces an event rate which is 90\% nuclear recoils) calibration sources at low and high voltages, inference of ultra-low-energy nuclear-recoil yield is possible.
	\item Fit systematics: the functional form for the parameterized fit also created the simulated Monte Carlo spectra. If actual backgrounds have different shapes, they may result in poor fits for both the low- and high-voltage rate distributions, leading to suppressed nuclear-recoil sensitivity.
 \end{itemize}
 
 \section{Conclusion}
With current CDMS technology and cryogenic backgrounds, phonon instrumented Ge detectors operated at high voltage have sensitivity to low mass WIMPs $\sim$2 orders of magnitude below the CoGeNT and DAMA signal regions. Furthermore, even exponentially increasing low-energy electronic backgrounds can be distinguished from a low-mass WIMP signal 
for cross sections  an order of magnitude smaller than the CoGeNT and DAMA signal regions by alternating between high- and low-voltage bias.

\begin{acknowledgements}
We would like to thank D. Moore and  J. Filippini for valuable discussions. This work is supported by the Department of Energy contract DE-FG02-04ER41295, and in part by the National Science Foundation Grant No. PHY-0855525.
\end{acknowledgements}


\end{document}